\documentclass[aps,10pt,pra,showpacs,nofootinbib,longbibliography,notitlepage]{revtex4-1}
\usepackage{hyperref}
\usepackage{amsmath}
\usepackage{amssymb}
\usepackage{amsthm}
\usepackage{graphics}
\usepackage{graphicx}
\usepackage{color}
\usepackage{soul}  
\usepackage[usenames,dvipsnames,svgnames]{xcolor}
\usepackage{lineno}
\usepackage{paralist}

\long\def\ca#1\cb{}

\def\Tr#1{\textrm{Tr}\left(#1\right)}

\def\AC{{\cal A}}

\def\CC{{\cal C}}

\def\EC{{\cal E}}

\def\HC{{\cal H}}

\def\KC{{\cal K}}
\def\LC{{\cal L}}

\def\SC{{\cal S}}

\def\endproof{{\hspace{\stretch{1}}$\blacksquare$}}

\def\eqref#1{{(\ref{#1})}}

\newtheorem{thm1}{Theorem}
\newtheorem{thm2}[thm1]{Theorem}
\newtheorem{thm3}[thm1]{Theorem}
\newtheorem{thm4}[thm1]{Theorem}

\newtheorem{cor1}[thm1]{Corollary}
\newtheorem{cor2}[thm1]{Corollary}
\newtheorem{lem2}[thm1]{Lemma}

\begin{document}
\title{Strong bounds on required resources for quantum channels by local operations and classical communication}
\author{Scott M. Cohen}
\email{cohensm52@gmail.com}
\affiliation{Department of Physics, Portland State University, Portland OR 97201}

\begin{abstract}
Given a protocol ${\cal P}$ that implements multipartite quantum channel ${\cal E}$ by repeated rounds of local operations and classical communication (LOCC), we construct an alternate LOCC protocol for ${\cal E}$ in no more rounds than ${\cal P}$ and no more than a fixed, constant number of outcomes for each local measurement, the same constant number for every party and every round. We then obtain another upper bound on the number of outcomes that, under certain conditions, improves on the first. The latter bound shows that for LOCC channels that are extreme points of the convex set of all quantum channels, the parties can restrict the number of outcomes in their individual local measurements to no more than the square of their local Hilbert space dimension, $d_\alpha$, suggesting a possible link between the required resources for LOCC and the convex structure of the set of all quantum channels. Our bounds on the number of outcomes indicating the need for only constant resources per round, independent of the number of rounds $r$ including when that number is infinite, are a stark contrast to the exponential $r$-dependence in the only previously published bound of which we are aware. If a lower bound is known on the number of product operators needed to represent the channel, we obtain a lower bound on the number of rounds required to implement the given channel by LOCC. Finally, we show that when the quantum channel is not required but only that a given task be implemented deterministically, then no more than $d_\alpha^2$ outcomes are needed for each local measurement by party $\alpha$.
\end{abstract}

\date{\today}

\maketitle
Economics is the study of how people use scarce resources to produce commodities for later consumption. Scientists designing experiments, as well as those who use their discoveries in everyday life, must take economics into account. These considerations have recently spawned studies of quantum resource theories \cite{HorodeckiOppenheimResource,ChitGourResource,BrandaoGourResource}, an outgrowth of quantum information science \cite{NielsenChuang}, which aim to quantify the costs of implementing exciting applications such as quantum teleportation \cite{BennettTele} and quantum computing \cite{Benioff,FeynmanQComp,Deutsch} under constraints imposed by circumstances. Entanglement \cite{HoroRMP}, a fascinating property found in multipartite quantum systems involving an unusual and extremely strong type of correlation between the parts of these systems, was the first example of such a resource theory. An example of constraints imposed is the recognition that entanglement cannot be created by Local Operations on constituent subsystems, even when supplemented by Classical Communication between the acting parties, a paradigm known as LOCC \cite{BennettConcentrate}, which plays an important role in many aspects of quantum information processing, including distributed quantum computing \cite{CiracDistComp}, entanglement distillation \cite{BennettPurifyTele} and manipulation \cite{Nielsen,BKrausStateTransf}, local distinguishability of quantum states \cite{Walgate}, local cloning \cite{Anselmi}, and various quantum cryptographic protocols, such as secret sharing \cite{HillerySecret}. Under the constraint of LOCC, which may arise due to spatial separation between constituent parts of a larger system, any entanglement must be supplied in advance and is therefore a valuable resource, whereas local operations and classical communication are each viewed as free. In reality, however, LOCC itself comes at a cost. Classical communication channels must be available to carry information from one party to another, those parties must have the means to make complex local measurements, and they must have the time available to implement the possibly numerous rounds of measuring and sharing information that may be needed.

The number of rounds needed for LOCC has received a fair amount of attention over recent years \cite{ChitambarHsiehRnd}, showing that there are circumstances when a single round is as good as many \cite{LoPopescu} whereas in other cases, two or more rounds \cite{BennettMixedQEC}, possibly even an infinite number \cite{Chitambar}, are necessary. In this paper, we consider a different resource for LOCC, a property that has heretofore received little attention: the complexity of making measurements and the corresponding amount of classical communication that must be exchanged between the parties at each round, both of which are determined by the number of outcomes in individual local measurements. Generalized measurements on system $\SC$ can be performed by introducing an ancillary system $\SC_a$ of dimension equal to the number of outcomes, interacting $\SC_a$ with $\SC$  and then performing a (projective) measurement on $\SC_a$. As recently shown in \cite{LloydTime4U}, the interaction between $\SC_a$ and $\SC$ requires a time scaling as the squared dimension of the combined system $\SC_a\otimes\SC$. In addition, the ability to distinguish among the various outcomes of the measurement---which may, for example, be made using a Stern-Gerlach \cite{SternGerlach1,SternGerlach2} type of apparatus---requires a level of spatial resolution that will also scale as the dimension of $\SC_a$. Thus, resource requirements increase with the number of outcomes of the desired measurement on $\SC$. The question of the number of outcomes necessary at each round has been previously addressed for the case of finite-round LOCC protocols, in the context of showing that the set of all such protocols is compact \cite{WinterLeung}. It was shown there that if quantum channel $\EC$ acts on a multipartite system of overall dimension $D$ and can be implemented by an LOCC protocol consisting of a total of $r$ rounds, then there exists an $r$-round protocol for $\EC$ such that at round $l$ there are no more than $D^{4(r-l+1)}$ outcomes in any local measurement \cite{WinterLeung}. This result was important in that prior to this work, there was every possibility that ``intermediate measurements with an unbounded number of outcomes" could be necessary at any point during the protocol \cite{WinterLeung}. The result scales poorly with $r$, however, and leaves open the possibility that an unbounded number of outcomes continue to be necessary within infinite-round protocols. Below, we obtain upper bounds on the required number of outcomes at each round which are (1) independent of $r$ (and of $l$), (2) never greater than $D^4$, and (3) applicable to both finite- and infinite-round protocols, strongly bounding the required resources for LOCC. Depending on certain parameters, these bounds can be surprisingly small.

Our upper bounds on the number of outcomes in local measurements can have important practical applications, examples of which have already been mentioned above. If, as had been believed prior to the present results, a protocol had required any given party to make measurements at different rounds with widely differing numbers of outcomes, they would have needed to either use different measuring apparatuses for different measurements or else use a single apparatus capable of measurements with a larger number of outcomes than necessary. The latter option would require larger ancillae and thus, more space and more time \cite{LloydTime4U} to implement than would at times be necessary, whereas the former option would obviously be more costly in a variety of ways. Therefore, the ability to limit the required number of outcomes at every round throughout the protocol can represent a significant savings in resources. It should also be mentioned that our upper bounds can be useful in simplifying the design of protocols \cite{myQChImpossbyLOCC}, making them more computationally tractable by reducing the number of outcomes that one must search for at each round.

After completing a draft of this manuscript, we learned that Leung, Winter and Yu \cite{LeungWinterYu} have recently arrived at results that are very similar in spirit to our own. They have also shown how to `compress' an LOCC protocol to one that has a limited number of outcomes for each local measurement, and their compression will often yield a tighter bound than ours, but the reverse situation can also hold. Their main results apply to finding optimal protocols for certain tasks, whereas we are here concerned with quantum channels. They also obtain a result for quantum channels, but our results have certain advantages over theirs in this case. First, their method is such that the compression of each local measurement depends on the entire protocol, whereas for our method each local measurement is compressed without reference to any other part of the protocol. In addition, their result requires that the parties use shared randomness, but our results do not require any randomness at all.

We consider the evolution, $\EC$, of a multipartite quantum system interacting with an environment, viewed as a noisy quantum channel \cite{LloydQCap,Schumacher} mapping initial quantum state $\rho$, an operator on Hilbert space $\HC$, to state $\EC(\rho)$. The channel $\EC$ may be represented in terms of a set of Kraus operators \cite{Kraus}, $K_i$, as
\begin{align}\label{eqn19}
\EC(\rho) = \sum_{i=1}^NK_i\rho K_i^\dag,
\end{align}
with
\begin{align}\label{eqn20}
\sum_{i=1}^NK_i^\dag K_i=I_\HC,
\end{align}
where $I_\HC$ is the identity operator on $\HC$. This collection of Kraus operators is referred to as a Kraus representation of $\EC$, and such representations are not unique. Without loss of generality, we assume that set $\{K_i\}$ is a minimal set, in the sense that no smaller set of Kraus operators represents $\EC$. The Kraus rank $\kappa$ of $\EC$ is defined as the size of this minimal set, so $N=\kappa$, where $1\le\kappa\le D^2$. Any other set of Kraus operators, $\{K_j^\prime\}_{j=1}^{N^\prime}$, describes the same channel as the original set if and only if there exists \cite{NielsenChuang} an isometry $V$, $V^\dag V=I_\kappa$ with  $I_\kappa$ the $\kappa\times\kappa$ identity matrix, such that for each $j=1,\cdots,N^\prime$,
\begin{align}\label{eqn21}
K^\prime_j=\sum_{i=1}^{\kappa}V_{ji}K_i.
\end{align}

We consider quantum channels implemented by LOCC protocols. It is common to represent an LOCC protocol as a rooted tree, which is a tree graph (no cycles) in which one node has been designated as the root and every edge is directed away from the root. The root represents the situation before the parties have done anything. Each node has a set of children, each child representing one of the outcomes of a measurement made at that stage of the protocol. A branch starts at the root and stretches via an edge from each node on that branch to one of its children, either continuing without end or else terminating at what is known as a leaf node. As we have discussed elsewhere \cite{myQChImpossbyLOCC}, if the tree $\LC$ represents an LOCC protocol (finite or infinite), which implements the set of Kraus operators $\{K_j^\prime\}$, then: (a) to each node $n\in\LC$ may be associated a positive semidefinite operator $E_n$ representing the accumulated action of all parties up to that node $n$; (b) $E_n$ is equal to a positive linear combination of the set of operators, $\{K_j^{\prime\dag}K_j^\prime\}$; and (c) each leaf node is proportional to one of the $K_j^{\prime\dag}K_j^\prime$ with positive constant of proportionality. Then, if this tree $\LC$ implements quantum channel $\EC$, we see using Eq.~\eqref{eqn21} that
\begin{align}\label{eqn22}
	E_n &= \sum_{j=1}^{N^\prime} c_j^{(n)} K_j^{\prime\dag} K_j^\prime = \sum_{i,i^\prime=1}^{\kappa}\left[\sum_{j=1}^{N^\prime}V_{ji}^\ast c_j^{(n)}V_{ji^\prime}\right]K_i^\dag K_{i^\prime}\notag\\
				&=\sum_{i,i^\prime=1}^{\kappa}\CC_{ii^\prime}^{(n)}K_i^\dag K_{i^\prime},
\end{align}
with $c_j^{(n)}\ge0$ for all $j,n$ and $\CC_{ii^\prime}^{(n)} := \sum_{j=1}^{N^\prime}V_{ji}^\ast c_j^{(n)}V_{ji^\prime}$. Since $V$ is an isometry, $\CC^{(n)}$ is a positive semidefinite matrix, $\CC^{(n)}\ge0$. Note also that if node $n$ is a leaf node, then as pointed out above, $E_n=c_J^{(n)}K_J^{\prime\dag}K_J^\prime$ for some fixed $J$ (no summation). This implies that the rank of matrix $\CC^{(n)}$ is equal to unity for each leaf node in the tree. Given that $\CC^{(n)}\ge0$, this means that $\CC^{(n)}=\vec{v_n}\vec{v_n}^{\dag}$ at a leaf node,  for some vector $\vec{v_n}\in\mathbb{C}^{\kappa}$. Specifically, we see that $\left(\vec{v_n}\right)_i=\sqrt{c_J^{(n)}}V_{Ji}^\ast$, and the Kraus operator associated with this leaf is $K_n^\prime=\sum_i\left(\vec{v_n}\right)_iK_i$.

The main tool for our arguments will be the following theorem. It provides a necessary and sufficient condition for a tree graph to represent an LOCC protocol implementing quantum channel $\EC$, with each node in the tree labeled by a positive semi-definite $\kappa\times\kappa$ matrix $\CC^{(n)}$.
\begin{thm1}\label{thm1}Quantum channel $\EC$, represented by the (minimal) set of Kraus operators $\{K_i\}_{i=1}^\kappa$, can be implemented by $\textrm{LOCC}$ if and only if there exists a tree graph $\LC$ satisfying all of the following conditions: 
\begin{enumerate}
\item\label{itm2} Each node $n\in\LC$ is labeled by a $\kappa\times \kappa$ matrix $\CC^{(n)}\ge0$ such that $\sum_{i,i^\prime}\CC^{(n)}_{ii^\prime}K_i^\dag K_{i^\prime}$ is a product operator;  
 \item\label{itm1} The root node is labeled by matrix $\CC^{(0)}=I_\kappa$, so that $\sum_{i,i^\prime}\CC^{(0)}_{ii^\prime}K_i^\dag K_{i^\prime}=I_\HC$;   
 \item\label{itm4} For each node $n$ and matrix $\CC^{(n)}$, the collection of its child nodes, $s$, with matrices $\CC^{(s)}$, satisfy $\sum_{s\in \textrm{siblings}}\CC^{(s)}=\CC^{(n)}$;
 \item\label{itm3} Each node $n$ along with each of its child nodes, $s\in\textrm{siblings}$, correspond to positive semidefinite product operators $E_n$ and $E_s$ that differ in only one party's local operator, that being the same party for all of them. For example, if it is party $A$, all sibling nodes are of the form $E_s=\sum_{i,i^\prime}\CC_{ii^\prime}^{(s)}K_i^\dag K_{i^\prime}=\AC^{(s)}\otimes\bar\AC$ with parent $E_n=\AC^{(n)}\otimes\bar\AC$, where $\bar\AC$ is a positive semidefinite (product) operator acting on all parties other than $A$ (the same operator for all of these nodes), and by the preceding item~\ref{itm4}, $\sum_{s\in\textrm{siblings}}\AC^{(s)}=\AC^{(n)}$;
  \item \label{itm5}\begin{enumerate}
 \item\label{itm5a}for every leaf node $l$, $\CC^{(l)}=\vec{v_l}\vec{v_l}^\dag$, and so has rank equal to one;
  \item\label{itm5b} for every infinite branch of $\LC$ (if any), the condition of the preceding item (\ref{itm5a}) is satisfied asymptotically;
\end{enumerate}
  \item \label{itm6} The sum of all leaf nodes, along with asymptotic approaches along infinite branches (with appropriate limiting procedure) is equal to the root node: $\sum_l\CC^{(l)}=\CC^{(0)}=I_\kappa$.
\end{enumerate}
\end{thm1}
\proof Item~\ref{itm2} is just the well-known condition that LOCC protocols can only implement product operators. Item~\ref{itm1} just says that at the beginning of the protocol, no party has done anything yet. Item~\ref{itm4} is the condition that each local measurement is complete, in the sense that the probabilities of all outcomes of any given measurement must sum to unity. Item~\ref{itm3} is the condition that the parties take turns making measurements, only one party measuring at any given time. Item~\ref{itm5} is necessary and sufficient that the collection of final outcomes of the protocol constitute an implementation of the channel $\EC$, a conclusion that follows from the discussion after Eq.~\eqref{eqn22}. Item~\ref{itm6} follows directly from repeated application of Item~\ref{itm4}.\endproof

Our main results will be direct consequences of the following lemma.
\begin{lem2}\label{lem2}
If multipartite quantum channel $\EC$ can be implemented by LOCC, then there exists an LOCC implementation of $\EC$ such that for each local measurement, the collection of matrices $\{\CC^{(s)}\}_s$ corresponding to the outcomes of that measurement (these outcomes indexed by $s$) is a linearly independent set.
\end{lem2}
\proof Let us consider node $n\in\LC$, and assume matrices $\CC^{(s)}$ associated with its children collectively form a linearly dependent set. We will give a constructive argument showing that the number of these children can be reduced by unity. This will prove the lemma, since we can continue this process for as long as the children remain dependent.

Since $\CC^{(s)}\ge0$, linear dependence implies the existence of a vanishing real linear combination of these matrices. By a judicious choice of $s_1$, there then exist coefficients $q_s$ such that
\begin{align}\label{eqn901}
\CC^{(s_1)}=\sum_{s\ne s_1}q_s\CC^{(s)},
\end{align}
with $\vert q_s\vert\le1$ and the sum is over all siblings of $s_1$. This means that if we omit node $s_1$, along with all of its descendants, and replace each sibling of $s_1$ by $\CC^{(s)}\rightarrow(1+q_s)\CC^{(s)}$, then the sum of the remaining children of node $n$ is still equal to $\CC^{(n)}$, as required. Note that since $1+q_s\ge0$, these replacement matrices continue to be positive semidefinite. Therefore, this new local measurement is still a valid one at this stage of the protocol. Nonetheless, there remain two questions that need be considered before we are done. First, do the descendants of those remaining children continue to constitute a valid protocol? They do not, as they were, but they can easily be altered to become valid, by replacing every node, $t$, descendant from child node $s$, by $\CC^{(t)}\rightarrow(1+q_s)\CC^{(t)}$. By doing so, every set of children of a given node continues to satisfy the conditions of Theorem~\ref{thm1}; in particular, conditions \ref{itm2} and \ref{itm4} that the nodes are positive matrices that sum to their parent and correspond to positive semidefinite product operators are still satisfied.

This demonstrates that following omission of $s_1$ and all its descendants, we are still left with a valid LOCC protocol. However, there is still the following crucial question: Does this new protocol continue to implement the same, desired quantum channel $\EC$? The answer is yes. Note that the leaf nodes remaining in the trimmed tree are a subset of the same leaf nodes that were in the original protocol, though some have been modified as just described. As noted above, see the paragraph following Eq.~\eqref{eqn22}, each remaining leaf node $l$ is $\CC^{(l)}=\vec{v_l}\vec{v_l}^{\dag}$ with $\left(\vec{v_l}\right)_i=\sqrt{\tilde c_L^{(l)}}V_{Li}^\ast$ for some fixed index $L$, which depends on $l$ ($\tilde c_L^{(l)}$ is a product of $c_L^{(l)}$ and possibly a factor of $1+q_s$). For every infinite branch remaining, the same condition holds asymptotically. Furthermore, by construction, this collection of leaf nodes, along with these asymptotic approaches of the infinite branches, continues to satisfy conditions~\ref{itm1} and \ref{itm6} of Theorem~\ref{thm1}; that is, since the root node is unchanged in our procedure,
\begin{align}\label{eqn902}
\sum_l\CC^{(l)} = \CC^{(0)}=I_\kappa,
\end{align}
and summation here is over the leaf nodes and asymptotic approaches along infinite branches (with appropriate limiting procedures) remaining in the pruned tree. By looking at matrix elements of this expression, this yields
\begin{align}\label{eqn903}
\sum_l v_{lj}v_{li}^\ast=\delta_{ij}.
\end{align}
We see that the pruned tree implements the collection of Kraus operators (see the discussion following Eq.~\eqref{eqn22})
\begin{align}\label{eqn904}
K_l^{\prime\prime}=\sum_{j=1}^\kappa v_{lj}K_j,
\end{align}
where according to Eq.~\eqref{eqn903} the collection of matrix elements $v_{lj}$ constitute an isometry.\footnote{What we have effectively done here is to delete a subset of rows in the original isometry $V$, scaling each remaining row by some non-negative factor. We have also essentially proven that for every isometry $V$ that corresponds to an LOCC protocol containing intermediate measurements involving sets of linearly `dependent children', what remains after this process of deletions and re-scalings is still an isometry.} By the isometric freedom in operator-sum representations of quantum channels (see, for example, p.~$370$ in \cite{NielsenChuang}), we thus see that this protocol implements a valid set of Kraus operators for the desired channel $\EC$, and our pruned tree does indeed represent a valid LOCC implementation of $\EC$. This completes the proof.\endproof

Recall from earlier discussion that matrices $\CC^{(s)}\ge0$ are of size $\kappa\times\kappa$ implying there can be no more than $\kappa^2$ of these matrices in any linearly independent set. In addition, the proof of the preceding lemma constructs a new LOCC protocol for $\EC$ in no more rounds than the original protocol. Therefore, the next theorem follows immediately from Lemma~\ref{lem2}.
\begin{thm2}\label{thm2}
If multipartite quantum channel $\EC$ can be implemented by LOCC in $r$ rounds, where $r$ may be infinite, then there exists an LOCC implementation of $\EC$ using no more than $r$ rounds such that no local measurement in the protocol has more than $\kappa^2$ outcomes.
\end{thm2}
\noindent This upper bound on the number of outcomes needed in intermediate measurements is independent of the size of the Kraus representation actually implemented, and it is even independent of the size, $N_p$, of the smallest possible product Kraus representation; it only depends on $\kappa$. If a channel is LOCC in a finite number of rounds, $r$, then after reducing the protocol as described above, there will be no more than $\kappa^{2r}$ leaf nodes in the tree representing the resulting protocol. Since each Kraus operator in the Kraus representation implemented by this finite-round protocol must be a leaf node, we have that $N_p\le\kappa^{2r}$. Therefore, the following corollary is a direct consequence of Theorem~\ref{thm2}.
\begin{cor2}\label{cor2}
If the smallest Kraus representation of quantum channel $\EC$ by product Kraus operators has at least $N_p$ members, then the number of rounds, $r$, required to implement $\EC$ by LOCC is lower bounded as $r\ge\log{N_p}/\log{\kappa^2}$.
\end{cor2}
\noindent Theorem~\ref{thm2} applies in complete generality to all quantum channels, and depends only on the Kraus rank $\kappa$ of the given channel. It is thus seen to be a strong result, indicating a significant constraint on required resources, especially when $\kappa$ is small. It turns out, however, that this result can often be strengthened, in some cases considerably. This stronger result is stated in the following theorem, where we denote as $\chi$ the dimension of the subspace spanned by the set of operators $\{K_i^\dag K_j\}_{i,j=1}^\kappa$, and it is not difficult to show that $\chi$ is a characteristic property of the channel, being independent of the chosen Kraus representation.
\begin{thm3}\label{thm3}
If multipartite quantum channel $\EC$ has Kraus rank $\kappa$ and can be implemented by LOCC in $r$ rounds, where $r$ may be infinite, then there exists an LOCC implementation of $\EC$ using no more than $r$ rounds such that no local measurement in the protocol has more than $d_\alpha^2 +\kappa^2-\chi$ outcomes, for each party $\alpha$ ($d_\alpha$ is the dimension of the Hilbert space describing the states of party $\alpha$'s subsystem).
\end{thm3}
\proof  If $\chi\le d_\alpha^2$, then the bound of Theorem~\ref{thm2} already implies this result, so assume $\chi>d_\alpha^2$. Let us consider a measurement by Alice (party $A$) consisting of $N_A$ outcomes $\AC_s\otimes\bar\AC$, where $\bar\AC$ is an operator acting on the composite of all subsystems other than $A$, this operator being the same for all outcomes because only Alice is measuring, see Item~\ref{itm3} of Theorem~\ref{thm1}. Each outcome is associated with a matrix $\CC^{(s)}$, so by Lemma~\ref{lem2}, proof of this theorem will follow from showing that no more than $d_A^2+\kappa^2-\chi$ of these matrices can be linearly independent. Let us write
\begin{align}\label{eqn906}
\CC^{(s)}=\sum_{t=1}^{\kappa^2}M_{ st}Q^{(t)},
\end{align}
where the $Q^{(t)}$ are an orthonormal basis of the space of $\kappa\times\kappa$ matrices, $\Tr{Q^{(t)\dag}Q^{(t^\prime)}}=\delta_{tt^\prime}$, and we also choose these basis elements such that they correspond to the $\kappa^2-\chi$ (independent) linear dependencies of the $K_i^\dag K_j$; that is,
\begin{align}\label{eqn907}
0=\sum_{i,j=1}^{\kappa}Q_{ij}^{(t)}K_i^\dag K_j,
\end{align}
when $t>\chi$. Then we can write the outcomes of Alice's measurement as
\begin{align}\label{eqn905}
\AC_s\otimes\bar\AC=\sum_{i,j=1}^{\kappa}\CC_{ij}^{(s)}K_i^\dag K_j=\sum_{t=1}^{\kappa^2}M_{ st}\sum_{i,j=1}^{\kappa}Q_{{ij}}^{(t)}K_i^\dag K_j=\sum_{t=1}^{\chi}M_{ st}\KC^{(t)},
\end{align}
with $\KC^{(t)}:=\sum_{ij} Q_{{ij}}^{(t)}K_i^\dag K_j$, and we have used Eq.~\eqref{eqn907} to reduce the sum to just $\chi$ terms in the final expression. Note that orthogonality of the $Q^{(t)}$ matrices ensures that operators $\KC^{(t)}$ are linearly independent for $t=1,\ldots,\chi$ because otherwise there would be additional dependencies among the $K_i^\dag K_j$, contrary to assumption.

Since $\chi\le\kappa^2$, if the number of outcomes satisfies $N_A\le d_A^2$, then $N_A\le d_A^2+\kappa^2-\chi$, which is what we are trying to prove. Therefore, we only need consider the case that $N_A>d_A^2$. So suppose there are $N_A>d_A^2$ outcomes $\AC_s\otimes\bar\AC$, $s=1,\ldots,N_A$. Then since with $\bar\AC$ fixed, no more than $d_A^2$ of these can be linearly independent, there must exist $N_A-d_A^2$ linearly independent vectors $\vec\lambda^{(p)}$ in $N_A$-dimensions having elements $\lambda_s^{(p)}$ not all zero such that
\begin{align}\label{eqn908}
0=\sum_{s=1}^{N_A}\lambda_s^{(p)\ast}\AC_s\otimes\bar\AC=\sum_{t=1}^{\chi}\sum_{s=1}^{N_A}\lambda_s^{(p)\ast}M_{ st}\KC^{(t)},
\end{align}
for $p=1,\ldots,N_A-d_A^2$. By the linear independence of the $\KC^{(t)}$ that appear in this expression, this implies that
\begin{align}\label{eqn909}
\sum_{s=1}^{N_A}\lambda_s^{(p)\ast}M_{ st}=0,
\end{align}
for all $t\le \chi$. This means that the first $\chi$ columns of matrix $M$, consisting of matrix elements $M_{ st}$, are each orthogonal to the $(N_A-d_A^2)$-dimensional subspace spanned by the collection of vectors $\vec\lambda^{(p)}$. Since these columns of $M$ are $N_A$-dimensional, there can be no more than $d_A^2$ of them in any linearly independent subset. There are $\kappa^2-\chi$ remaining columns in $M$, so the rank of $M$ cannot exceed $d_A^2+\kappa^2-\chi$. This implies that the entire collection of columns of $M$ span a subspace of dimension no more than $d_A^2+\kappa^2-\chi$. Therefore, if $N_A\ge d_A^2+\kappa^2-\chi$, there exists a non-zero $N_A$-dimensional vector $\vec\lambda^\prime$, elements $\lambda_s^\prime$, orthogonal to every column of $M$. That is, $\sum_s \lambda_s^{\prime\ast} M_{ st}=0$ for all $t$. Multiplying Eq.~\eqref{eqn906} by $\lambda_s^{\prime\ast}$ and summing over $s$, we immediately see that the $\CC^{(s)}$ are linearly dependent, and we reach the conclusion that no more than $d_A^2+\kappa^2-\chi$ of the $\CC^{(s)}$ can be linearly independent. Since the same argument holds for any of the parties, and by reference to Lemma~\ref{lem2}, this completes the proof.\endproof

Finally, we note that channel $\EC$ is an extreme point of the convex set of all quantum channels if and only if $\chi=\kappa^2$ \cite{Choi}. This leads us to the following corollary as an immediate consequence of Theorem~\ref{thm3}, suggesting a possible connection between the required resources for LOCC and the convex structure of the set of all quantum channels.
\begin{cor1}\label{cor1}
If $\EC$ is an extreme point of the set of all quantum channels and can be implemented by LOCC in $r$ rounds, where $r$ may be infinite, then it can be implemented by an LOCC protocol using no more than $r$ rounds in which, for each party $\alpha$, no more than $d_\alpha^2$ outcomes are used in each local measurement by that party in the entire sequence of rounds of the protocol.
\end{cor1}

In summary, we have derived strong upper bounds on the number of outcomes that each intermediate measurement need have in any LOCC protocol implementing a quantum channel $\EC$, no matter how many rounds are involved, including in the limit of infinite rounds. These bounds are presented in Theorems~\ref{thm2} and \ref{thm3}, and they are independent of the round number and of the total number of rounds. These theorems lead directly to Corollary~\ref{cor2}, which provides a lower bound on the number of rounds needed to implement $\EC$ if a lower bound on the number of product operators in any product Kraus representation of $\EC$ is known, and Corollary \ref{cor1}, which suggests a possible link between the convex structure of the set of quantum channels and the resources needed to implement those channels by LOCC. It is perhaps worth pointing out that Theorem~\ref{thm2} (and thus, Corollary~\ref{cor2}), which depends only on the Kraus rank $\kappa$ of $\EC$, provides a bound that is independent of the way the parties are partitioned. That is, for example, if two or more of the original parties are able to merge together to act as one, $\kappa$, and therefore this bound, remains unchanged. Note also that $\chi$ is independent of partitioning, as well. Therefore, merging (or splitting) the parties only changes the bound in Theorem~\ref{thm3} by changing the local dimensions $d_\alpha$.

We would like to make clear that the crucial step in achieving these results is the representation of LOCC protocols in terms of the $\kappa\times\kappa$ matrices $\CC^{(n)}$, introduced in \cite{myQChImpossbyLOCC} and described here in Eq.~\eqref{eqn22}. It is these matrices that have allowed us to show that our pruning procedure, described in the proof of the critical Lemma~\ref{lem2} where all sets of sibling nodes are reduced to linearly independent sets of these $\CC^{(n)}$ matrices, leaves the quantum channel implemented by the protocol unchanged; see the argument leading to Eqs.~\eqref{eqn903} and \eqref{eqn904}. One might try to prune these trees differently, in ways that leave sets of sibling nodes linearly independent when viewing these nodes as positive semidefinite operators $E_n$, or even as Kraus operators, but such efforts appear to be doomed to failure. In the case of labeling by $E_n$, one can easily prune in this way such that the tree remains a valid LOCC protocol, but it will not generally implement the same channel as the unpruned tree. For example, suppose in the original protocol for $\EC$ there is a terminal measurement (every outcome corresponding to a leaf node) where the collection of sibling operators $E_s$ obeys a single linear dependency, but that of matrices $\CC^{(s)}$---which have rank equal to unity according to Item~\ref{itm5} of Theorem~\ref{thm1}---is linearly independent. Then, in eliminating one of the $E_s$ (say $E_{s_1}$, with $s_1$ one of the original leaf nodes) to end up with a linearly independent set of the remaining $E_s$, the remaining $\CC^{(s)}$ matrices will be replaced by new matrices that are positive linear combinations of two of the original $\CC^{(s)}$ matrices; specifically, $\CC^{(s)}+q_{s}\CC^{(s_1)}$. Since $\CC^{(s)}$ and $\CC^{(s_1)}$ are linearly independent, rank-$1$, positive definite matrices, such linear combinations cannot themselves have rank equal to unity and thus, again according to Item~\ref{itm5} of Theorem~\ref{thm1}, the resulting LOCC protocol no longer implements the original channel $\EC$. Of course, this raises the question of whether it may be possible to improve on these bounds, or if our bounds may themselves be tight for the implementation of a given quantum channel (while preserving the top-down nature of the approach, or without the availability of sufficient shared randomness \cite{LeungWinterYu}). We are presently unable to provide an intelligent guess as to which of these possibilities is more likely, but it is certainly a question that deserves further study.

Notice that pruning in a way that leaves sibling $E_s$ operators linearly independent fails only because of the need to retain the rank-$1$ condition on leaves of the tree in order to preserve implementation of the same channel $\EC$. It may be that we do not need to implement a given channel but instead only care about accomplishing a given task. Consider, then, a `deterministic' LOCC protocol, by which we mean that every leaf in the tree is successful in completing a given task, and for each infinite branch (if any), successful completion of the task is approached asymptotically. Examples include deterministic local state transformation \cite {Nielsen,BKrausStateTransf} and local cloning \cite{ourLocalCloning}. Since every branch in such protocols is successful, we may prune these trees to leave sibling $E_s$ operators linearly independent such that the resulting tree---whose branches are a subset of those in the original protocol---also represents a deterministic protocol for the given task. Since each local measurement (for example, by Alice) produces sibling $E_s$ operators of the form $\AC_s\otimes\overline\AC$ with $\overline\AC$ the same for all these children, and since the $\AC_s$ are $d_A\times d_A$ matrices, we have thus proved, and will end with, the following theorem.
\begin{thm4}\label{thm4}
Suppose we have an LOCC protocol that deterministically implements a given task in $r$ rounds, where $r$ may be infinite. Then there exists an LOCC protocol that also deterministically implements the given task using no more than $r$ rounds, and such that no local measurement in the protocol has more than $d_\alpha^2$ outcomes for each party $\alpha$.
\end{thm4}

\noindent\textit{Acknowledgments} --- We wish to thank Debbie Leung for helpful discussions, and Barbara Kraus for raising the question about deterministic protocols addressed in our Theorem~\ref{thm4}.


%

\end{document}